\documentclass[10pt,a4paper,tbtags, sort&compress]{article}

\usepackage{amsmath}
\usepackage{amssymb}
\usepackage{mathtools}
\usepackage{mathrsfs}
\usepackage{url}
\usepackage{enumitem}
\usepackage[square,sort,comma,numbers]{natbib}
\usepackage{gensymb}
\usepackage{caption}
\DeclareMathOperator{\sech}{sech}
\DeclareMathOperator{\sgn}{sgn}

\begin{document}
\title{Multi-Particle Invariant Mass \textemdash\ Standard Expressions and Corrections to Order $(m/E)^4$}
\author{M.P. Fewell \cr ARC Centre of Excellence for Dark-Matter Particle Physics \cr and \cr Department of Physics, Adelaide University, \cr Adelaide, South Australia 5005, Australia}
\date{February 2026}
\maketitle
\begin{abstract} \noindent
In collider-based particle physics, \emph{invariant mass} refers to the magnitude of the total-momentum 4-vector of a system of particles. An expression for the invariant mass of a 2-particle system is well known; it assumes that both the total energy $E$ and the transverse momentum $p_\mathrm{T}$ of each particle in the system greatly exceed its mass $m$. This note explores these assumptions by computing correction terms in powers of $m/E$ up to order $(m/E)^4$. The assumptions are found to be robust: not only is the leading correction quadratic in $m/E$, but also cancellations reduce its coefficient and that of the next-to-leading correction, which is of order $(m/E)^4$. Three- and four-particle systems are also treated and the generalisation to larger numbers of particles indicated. The zeroth-order expressions for these multi-particle systems are remarkably simple; they deserve to be better known.
\end{abstract}

\section{Introduction \textemdash\, basics}
\noindent The Wikipedia page on invariant mass \citep{Wi25} and its associated discussion page leave one with the impression that the term is ill-defined and in dispute. By contrast, in collider-based particle physics the definition of the invariant mass $m$ of a system of particles is clear and universally accepted: it is the magnitude of the total-momentum 4-vector of the particles (e.g. \citep[p. 1-8]{Bu21}):\footnote[1]{In particle physics, the quantity denoted $m^2$ herein is more conventionally labelled $s$.}
\begin{equation} \label{invMdef}
    m^2 = \sum_\mu \left[ \sum_i p^\mu_i \; \sum_j p_{\mu j} \right],
\end{equation}
\noindent where $p^\mu_i$ is the momentum 4-vector of particle $i$ and the sum over $\mu$ is the Einstein summation convention explicitly written out. It is useful for two reasons: because the magnitude of a 4-vector is invariant under Lorentz transformations and because of the centrality of momentum conservation in particle interactions.

The 4-momentum of a single particle has contravariant and covariant components
\begin{equation*}
    p^\mu = \left[ E, p_\mathrm{x}, p_\mathrm{y}, p_\mathrm{z} \right], \qquad p_\mu = \left[ E, -p_\mathrm{x}, -p_\mathrm{y}, -p_\mathrm{z} \right]
\end{equation*}
\noindent respectively, where $E$ is the total energy of the particle and $p_\mathrm{x}$ etc. are the components of its momentum 3-vector $\mathbf{p}$. This motivates naming the quantity in eq. \eqref{invMdef} as a ``mass'', owing to the fundamental relation\footnote[2]{This note uses natural units throughout, in which the speed $c$ of light is set to unity.}
\begin{equation*}
    m^2 =  E^2 - \mathbf{p}^2
\end{equation*}

The standard derivation of invariant-mass expressions in accelerator-based particle physics makes the ultra-relativistic assumption $E \gg m$, with $m$ neglected, for each particle in the system. This note reports the derivation of correction terms in $m/E$. Section \ref{sec:basics} sets the scene by recapitulating the standard derivation of the invariant mass of two to four particles, from which the generalisation to more particles is straightforwardly apparent. Section \ref{sec:corr} derives correction terms, which turn out to involve powers of $(m/E)^2$.

\section{Standard expressions}
\label {sec:basics}
\noindent By convention, one aligns the z axis with the direction of a Lorentz boost \textemdash\ the beam direction in a collider experiment \textemdash\ so the azimuthal angle $\phi$ is unchanged by the Lorentz transformation, as is the transverse momentum $p_\mathrm{T}$.\footnote[3]{$p^2_\mathrm{T} = \mathbf{p}^2 - p^2_\mathrm{z}$} Hence
\begin{equation} \label{pxy}
    p_\mathrm{x} = p_\mathrm{T}\cos\phi, \qquad p_\mathrm{y} = p_\mathrm{T}\sin\phi.
\end{equation}

Since the polar angle $\theta$ \textbf{is} altered by the Lorentz transformation, it is usual to replace it with the rapidity $y$, defined by
\begin{equation} \label{rapidityDef}
    y = \frac{1}{2} \ln\left(\frac{E + p_\mathrm{z}}{E - p_\mathrm{z}}\right);
\end{equation}
\noindent for rapidity differences are Lorentz invariant (e.g. \citep[p. 1-17]{Bu21}). A little manipulation shows that
\begin{equation} \label{Epz}
    E = \sqrt{p^2_{\mathrm{T}} + m^2}\cosh y, \qquad  p_\mathrm{z} = \sqrt{p^2_{\mathrm{T}} + m^2}\sinh y.
\end{equation}

The relationship between $y$ and $\theta$ can be written in closed form when $E \gg m$, for then \citep[p. 1-18]{Bu21}
\vspace{-0.15cm}
\begin{equation*}
    y \approx \eta = \frac{1}{2}\ln\frac{1+\cos\theta}{1-\cos\theta} = \mathrel{-}\ln\tan(\theta/2),
\end{equation*}
\noindent where $\eta$ is known as pseudorapidity. A simpler relation, useful in \S\ref{sec:corr}, is
\vspace{-0.15cm}
\begin{equation} \label{tanhEta}
    \tanh\eta = \cos\theta .
\end{equation}

The standard procedure for introducing the experimental quantities $p_{\mathrm{T}}$, $\theta$ and $\phi$ into eq. \eqref{invMdef} uses pseudorapidity $\eta$ in place of $\theta$, with the attendant assumption of $E\gg m$. The next subsection recapitulates the standard derivation for two particles, with \S\ref{sec:m1234} treating 3- and 4-particle systems.

\subsection{Two particles}
\label{sec:twoPart}
\noindent For two particles, eq. \eqref{invMdef} is
\vspace{-0.15cm}
\begin{equation} \label{m12}
    \begin{split}
        m^2_{12} &= \sum_\mu (p^\mu_1 + p^\mu_2)(p_{\mu 1} + p_{\mu 2})  \\
        &= (E_1+E_2)^2-[(p_{\mathrm{x}1}+p_{\mathrm{x}2})^2 + (p_{\mathrm{y}1}+p_{\mathrm{y}2})^2 + (p_{\mathrm{z}1}+p_{\mathrm{z}2})^2].
    \end{split}
\end{equation}

\noindent The terms involving $p_\mathrm{x}$ and $p_\mathrm{y}$ are reduced by applying eq. \eqref{pxy}:
\vspace{-0.20cm}
\begin{multline*}
    (p_{\mathrm{x}1}+p_{\mathrm{x}2})^2 + (p_{\mathrm{y}1}+p_{\mathrm{y}2})^2 = (p_{\mathrm{T}1}\cos\phi_1 + p_{\mathrm{T}2} \cos\phi_2)^2 \\ + (p_{\mathrm{T}1} \sin\phi_1 + p_{\mathrm{T}2}\sin\phi_2)^2
\end{multline*}
\vspace{-0.40cm}
\begin{equation} \label{pxySq} \begin{split}
    &= p_{\mathrm{T}1}^2 + p_{\mathrm{T}2}^2 + 2p_{\mathrm{T}1}p_{\mathrm{T}2} (\cos\phi_1\cos\phi_2 + \sin\phi_1\sin\phi_2) \\[0.10cm]
    &= p_{\mathrm{T}1}^2 + p_{\mathrm{T}2}^2 + 2p_{\mathrm{T}1}p_{\mathrm{T}2}\cos(\phi_1-\phi_2).
    \end{split}
\end{equation}

The terms in eq. \eqref{m12} involving $E$ and $p_\mathrm{z}$ are treated similarly, using eq. \eqref{Epz}. Here, the assumption of $E \gg m$ allows the replacement of $y$ with $\eta$, but one must also assume $p_\mathrm{T} \gg m$, to remove the explicit $m$ in eq. \eqref{Epz}. These two assumptions are related but not identical: the second fails for momenta directed sufficiently close to the z axis. That is, one expects correction terms to involve $\eta$ as well as $m/E$. With these two assumptions, one has
\vspace{-0.20cm}
\begin{multline*}
    (E_1+E_2)^2 - (p_{\mathrm{z}1}+p_{\mathrm{z}2})^2 = (p_{\mathrm{T}1}\cosh\eta_1 + p_{\mathrm{T}2}\cosh\eta_2)^2 \\ - (p_{\mathrm{T}1}\sinh\eta_1 + p_{\mathrm{T}2}\sinh\eta_2)^2
\end{multline*}
\vspace{-0.40cm}
\begin{equation*} \begin{split}
    &= p_{\mathrm{T}1}^2 + p_{\mathrm{T}2}^2 + 2p_{\mathrm{T}1}p_{\mathrm{T}2} (\cosh\eta_1\cosh\eta_2 - \sinh\eta_1\sinh\eta_2) \\[0.10cm]
    &= p_{\mathrm{T}1}^2 + p_{\mathrm{T}2}^2 + 2p_{\mathrm{T}1}p_{\mathrm{T}2}\cosh(\eta_1-\eta_2). \\[0.05cm]
\end{split} \end{equation*}
\noindent Insertion of this and eq. \eqref{pxySq} into eq. \eqref{m12} immediately gives the oft-quoted result (e.g. \citep{Wi25, Fi24, Di18, Fa12}):
\vspace{-0.10cm}
\begin{equation} \label{pairwiseM}
    m^2_{12} = 2p_{\mathrm{T}1}\,p_{\mathrm{T}2}\,[\cosh(\eta_1-\eta_2) - \cos(\phi_1-\phi_2)].
    \end{equation}

\subsection{Three and four particles, and more} \label{sec:m1234}
\noindent From eq. \eqref{invMdef}, the invariant mass of a system of three particles is
\begin{equation}
    m^2_{123} = \sum_\mu (p^\mu_1 + p^\mu_2  + p^\mu_3)(p_{\mu 1} + p_{\mu 2} + p_{\mu 3}).
\end{equation}
\vspace{-0.30cm}
\noindent Expanding this out gives
\begin{multline*}
    m^2_{123} = (E_1+E_2+E_3)^2-(p_{\mathrm{x}1}+p_{\mathrm{x}2}+p_{\mathrm{x}3})^2 \\ - (p_{\mathrm{y}1}+p_{\mathrm{y}2}+p_{\mathrm{y}3})^2 - (p_{\mathrm{z}1}+p_{\mathrm{z}2}+p_{\mathrm{z}3})^2.
\end{multline*}
\noindent Applying the approximations $E \gg m$ and $p_\mathrm{T} \gg m$:
\vspace{-0.15cm}
\begin{align*}
    m^2_{123} = (&p_{\mathrm{T}1}\cosh\eta_1 + p_{\mathrm{T}2}\cosh\eta_2 + p_{\mathrm{T}3} \cosh\eta_3)^2 \\ 
    &- (p_{\mathrm{T}1}\sinh\eta_1 + p_{\mathrm{T}2}\sinh\eta_2 + p_{\mathrm{T}3}\sinh\eta_3)^2 \\
    &- (p_{\mathrm{T}1}\cos\phi_1 + p_{\mathrm{T}2}\cos\phi_2 + p_{\mathrm{T}3}\cos\phi_3)^2 \\
    &- (p_{\mathrm{T}1}\sin\phi_1 + p_{\mathrm{T}2}\sin\phi_2 + p_{\mathrm{T}3}\sin\phi_3)^2,
\end{align*}
\noindent which, on multiplying out, reduces to
\vspace{-0.10cm}
\begin{equation} \label{tripletM}
    m^2_{123} = m^2_{12} + m^2_{13} + m^2_{23}\, .
\end{equation}
    
A similar analysis gives the square of the invariant mass of a system of four particles as the sum of the squared invariant masses of the pairwise combinations:
\vspace{-0.10cm}
\begin{equation} \label{quadrupletM}
    m^2_{1234} = m^2_{12} + m^2_{13} + m^2_{14} + m^2_{23} + m^2_{24} + m^2_{34}\, .
\end{equation}
\noindent The generalisation to five and more particles is clear.

\section{Corrections for the approximation $E \gg m$} \label{sec:corr}
\noindent The corrections to eqs \eqref{pairwiseM}, \eqref{tripletM} and \eqref{quadrupletM} are derived by expanding eqs \eqref{rapidityDef} and \eqref{Epz} in powers of $m/E$. Correction terms come from two sources: the use of pseudorapidity $\eta$ in place of true rapidity $y$ and the surd in eq. \eqref{Epz}.

\subsection{Use of pseudorapidity}
\noindent Rewriting eq. \eqref{rapidityDef} into a form suitable for the application of Taylor's theorem:
\begin{equation*} \begin{split}
    y &= \frac{1}{2} \ln\left(\frac{E + \sqrt{(E^2 - m^2)}\cos\theta}{E - \sqrt{(E^2 - m^2)}\cos\theta}\right) \\[0.15cm]
    &= \frac{1}{2} \ln\left(\frac{\sec\theta + \sqrt{1 - (m/E)^2}}{\sec\theta - \sqrt{1 - (m/E)^2}}\right).
\end{split} \end{equation*}
\noindent For $m/E$ small, this is, to second order in $(m/E)^2$,\footnote[4]{Here and below, an approximately-equal sign is used to indicate that terms on the right-hand side of order higher than $(m/E)^4$ have been dropped on this particular line.}
\begin{equation*} \begin{split}
    y &\approx \frac{1}{2} \ln\left(\frac{\sec\theta + 1 - (1/2)(m/E)^2 - (1/8)(m/E)^4}{\sec\theta - 1 + (1/2) (m/E)^2 + (1/8)(m/E)^4}\right) \\[0.1cm]
    &= \frac{1}{2} \ln\left(\frac{1 + \cos\theta - (1/2)(m/E)^2\cos\theta - (1/8)(m/E)^4\cos\theta}{1 - \cos\theta + (1/2) (m/E)^2\cos\theta + (1/8)(m/E)^4\cos\theta}\right) \\[0.1cm]
    &= \frac{1}{2} \ln \left\{\frac{1 + \cos\theta}{1 - \cos\theta}\left[1 - \frac{1}{2}\left(\frac{m}{E}\right)^2\frac{\cos\theta}{1 + \cos\theta} - \frac{1}{8}\left(\frac{m}{E}\right)^4\frac{\cos\theta}{1 + \cos\theta}\right] \right. \\ 
    &\left.\mspace{118mu} \times \left[1 + \frac{1}{2}\left(\frac{m}{E}\right)^2\frac{\cos\theta}{1 - \cos\theta} + \frac{1}{8}\left(\frac{m}{E}\right)^4\frac{\cos\theta}{1 - \cos\theta}\right]^{-1}\right\}\\[0.1cm]
    &\approx \frac{1}{2} \ln \left\{\frac{1 + \cos\theta}{1 - \cos\theta}\left[1 - \frac{1}{2}\left(\frac{m}{E}\right)^2\frac{\cos\theta}{1 + \cos\theta} - \frac{1}{8}\left(\frac{m}{E}\right)^4\frac{\cos\theta}{1 + \cos\theta}\right] \right. \\
    &\left.\mspace{118mu} \times \left[1 - \frac{1}{2}\left(\frac{m}{E}\right)^2 \frac{\cos\theta}{1 - \cos\theta} + \frac{1}{4}\left(\frac{m}{E}\right)^4 \left(\frac{\cos\theta}{1 - \cos\theta}\right)^2 \right.\right. \\
    &\left.\left.\mspace{162mu} \mathrel{-} \frac{1}{8}\left(\frac{m}{E}\right)^4\frac{\cos\theta}{1 - \cos\theta}\right]\right\} \\[0.10cm]
    &\approx \frac{1}{2} \ln \left\{\frac{1 + \cos\theta}{1 - \cos\theta} \left[1 - \left(\frac{m}{E}\right)^2\frac{\cos\theta}{\sin^2\theta} - \frac{1}{4} \left(\frac{m}{E}\right)^4 \frac{\cos\theta (1 - 3\cos\theta)}{\sin^2\theta (1 - \cos\theta)} \right]\right\} .
\end{split} \end{equation*}
Hence, to order $(m/E)^4$,
\begin{equation*} \begin{split}
    y &= \eta + \frac{1}{2} \ln \left[1 - \left(\frac{m}{E}\right)^2 \frac{\cos\theta}{\sin^2\theta} - \frac{1}{4}\left(\frac{m}{E}\right)^4 \frac{\cos\theta(1 - 3\cos\theta)}{\sin^2\theta(1 - \cos\theta)} \right] \\[0.1cm]
    &\approx \eta - \frac{1}{2}\left(\frac{m}{E}\right)^2\frac{\cos\theta}{\sin^2\theta} - \frac{1}{8}\left(\frac{m}{E}\right)^4 \frac{\cos\theta(1 - 3\cos\theta)}{\sin^2\theta(1 -\cos\theta)} - \frac{1}{4}\left(\frac{m}{E}\right)^4\frac{\cos^2\theta}{\sin^4\theta}
\end{split} \end{equation*}
\begin{equation} \label{yExpanded}
    \mspace{12mu} = \eta - \left(\frac{m}{E}\right)^2\frac{\cos\theta}{2\sin^2\theta} -\left(\frac{m}{E}\right)^4 \frac{\cos\theta(1 - 4\cos\theta - 3\cos^2\theta)}{8\sin^4\theta}. \mspace{41mu}
\end{equation} \vspace{0.10cm}

Equation \eqref{Epz} requires $\cosh y$ and $\sinh y$, so use
\vspace{-0.15cm}
\begin{equation*} \begin{split}
    \cosh(\eta + \alpha) &= \cosh\eta \cosh\alpha + \sinh\eta \sinh\alpha \\
     &\approx \left(1 + \alpha^2/2 \right) \cosh\eta + \alpha \sinh\eta
\end{split} \end{equation*}
\vspace{-0.10cm}
\noindent to second order in $\alpha$ for small $\alpha$. Similarly
\begin{equation*}
    \sinh (\eta +\alpha) \approx \left(1 + \alpha^2/2 \right) \sinh\eta + \alpha \cosh\eta.
\end{equation*}
\noindent Inserting eq. \eqref{yExpanded} into these leads to
\vspace{-0.10cm}
\begin{equation} \label{coshY} \begin{split}
    \cosh y &\approx \left[1 + \left(\frac{m}{E}\right)^4\frac{\cos^2\theta}{8\sin^4\theta} \right] \cosh\eta - \left[\left(\frac{m}{E}\right)^2 \frac{\cos\theta}{2\sin^2\theta} \right. \\[0.1cm]
    &\left.\qquad\;\; \mathrel{+} \left(\frac{m}{E}\right)^4  \frac{\cos\theta(1 - 4\cos\theta - 3\cos^2\theta)}{8\sin^4\theta}\right] \sinh\eta
\end{split} \end{equation}
\vspace{-0.25cm}
\noindent and similarly for $\sinh y$.

\vspace{0.35cm}
\subsection{Expanding the surd} \label{sec:surd}
\noindent The surd in eq. \eqref{Epz} is
\vspace{-0.20cm}
\begin{equation*}
    \sqrt{p^2_{\mathrm{T}} + m^2} = p_{\mathrm{T}} \sqrt{1 + \left(\frac{m}{p_{\mathrm{T}}}\right)^2}.
\end{equation*}
\noindent Now
\begin{equation*}
    p_{\mathrm{T}} = p \sin\theta = \sqrt{E^2 - m^2} \sin\theta.
\end{equation*}
\vspace{0.30cm}
\noindent Hence, in a form suitable for the application of Taylor's theorem,
\vspace{-0.15cm}
\begin{equation*} \begin{split}
    \sqrt{p^2_{\mathrm{T}} + m^2} &= p_{\mathrm{T}} \sqrt{1 + \frac{(m/E)^2}{\sin^2\theta\,[1 - (m/E)^2]}} \\[0.20cm]
    &\approx p_{\mathrm{T}} \sqrt{1 + \frac{m^2}{E^2\sin^2\theta}\left[1 + 
    \left(\frac{m}{E}\right)^2 + \left(\frac{m}{E}\right)^4 \right]} \\[0.20cm]
    &\approx p_{\mathrm{T}} \sqrt{1 + \left(\frac{m}{E\sin\theta}\right)^2 +\frac{m^4}{E^4\sin^2\theta}} \, .
\end{split} \end{equation*}
\noindent The involvement of $\theta$ is explicit here: one must assume that $\theta$ is sufficiently far from either zero or $\pi$. With this proviso,
\begin{equation} \label{surd} \begin{split}
    \sqrt{p^2_{\mathrm{T}} + m^2} &\approx p_{\mathrm{T}} \left[1 + \frac{1}{2} \left(\frac{m}{E\sin\theta}\right)^2 + \frac{1}{2}\frac{m^4}{E^4\sin^2\theta} - \frac{1}{8} \left(\frac{m}{E\sin\theta} \right)^4 \right] \\[0.15cm]
    &\approx p_{\mathrm{T}} \left[1 + \frac{1}{2} \left(\frac{m}{E\sin\theta} \right)^2 + \left(\frac{m}{E}\right)^4 \;\frac{ 4\sin^2\theta -1}{8\sin^4\theta} \right].
\end{split} \end{equation}
\noindent In fact, a similar $\theta$ dependence appears in eq. \eqref{coshY} \textemdash\, the $\sin\theta$ factors in the denominators. It is notable that both correction terms in eq. \eqref{yExpanded} are negative whereas both are positive in eq. \eqref{surd}. It reduces the magnitude of the corrections to $E$ and $p_\mathrm{z}$, as shown in the next subsection.

\subsection{Constructing $E$ and $p_\mathrm{z}$}
\noindent Inserting eqs \eqref{coshY} and \eqref{surd} into eq. \eqref{Epz}:
\begin{equation*} \begin{split}
    E &\approx p_{\mathrm{T}} \cosh\eta \left\{\left[1 + \frac{1}{2} \left(\frac{m}{E\sin\theta}\right)^2 + \left(\frac{m}{E}\right)^4 \;\frac{4 \sin^2\theta -1}{8\sin^4\theta}\right] \right. \\
    &\left.\mspace{105mu} \times \left[1 + \left(\frac{m}{E}\right)^4 \;\frac{\cos^2\theta}{8\sin^4\theta} - \left(\frac{m}{E}\right)^2 \frac{\cos\theta}{2\sin^2\theta}\tanh\eta  \right.\right. \\[0.1cm]
    &\left.\left.\mspace{144mu} + \left(\frac{m}{E}\right)^4 \;\frac{\cos\theta(1 - 4\cos\theta - 3\cos^2\theta)}{8\sin^4\theta} \tanh\eta \right]\right\} \\[0.25cm]
    &\approx p_{\mathrm{T}} \cosh\eta \left\{1 + \left(\frac{m}{E}\right)^2 \;\frac{1 - \cos\theta\tanh\eta}{2\sin^2\theta} + \left(\frac{m}{E}\right)^4 \left[\frac{4\sin^2\theta-1}{8\sin^4\theta} + \frac{\cos^2\theta}{8\sin^4\theta} \right.\right. \\[0.1cm]
    &\left.\left.\mspace{185mu} \mathrel{-} \frac{\cos\theta \tanh\eta(1 - 4\cos\theta - 3\cos^2\theta}{8\sin^4\theta} - \frac{\cos\theta\tanh\eta}{4\sin^4\theta} \right] \right\} \\[0.2cm]
    &= p_{\mathrm{T}} \cosh\eta \left[1 + \frac{1}{2}\left(\frac{m}{E}\right)^2 + \frac{1}{8} \left(\frac{m}{E}\right)^4 \;\frac{3 - 6\cos^2\theta + 4\cos^3\theta + 3\cos^4\theta}{\sin^4\theta} \right]
\end{split} \end{equation*}
\begin{equation} \label{Eexpanded}
 = p_{\mathrm{T}} \cosh\eta \left[1 + \frac{1}{2}\left(\frac{m}{E}\right)^2 + \frac{1}{8} \left(\frac{m}{E}\right)^4 \,\left(3 +\frac{ 4\cos^3\theta}{\sin^4\theta} \right) \right], \mspace{120mu}
\end{equation}
\noindent where the remarkable simplification in the last two lines \textemdash\ the leading correction has no $\theta$ dependence \textemdash\ comes from the use of eq. \eqref{tanhEta} in the previous line.

Similarly for $p_\mathrm{z}$:
\vspace{-0.10cm}
\begin{equation*} \begin{split}
    p_\mathrm{z} &\approx p_{\mathrm{T}} \sinh\eta \left\{\left[1 + \frac{1}{2} \left(\frac{m}{E\sin\theta}\right)^2 + \left(\frac{m}{E}\right)^4 \;\frac{4 \sin^2\theta -1}{8\sin^4\theta}\right] \right. \\[0.1cm]
    &\left.\mspace{105mu} \times \left[1 + \left(\frac{m}{E}\right)^4 \frac {\cos^2\theta}{8\sin^4\theta} - \left(\frac{m}{E}\right)^2 \frac{\cos\theta}{2\sin^2\theta}\coth\eta \right.\right. \\[0.1cm]
    &\left.\left.\mspace{145mu} + \left(\frac{m}{E}\right)^4 \;\frac{\cos\theta(1 - 4\cos\theta - 3\cos^2\theta)}{8\sin^4\theta} \coth\eta \right]\right\}.
\end{split} \end{equation*}
\noindent This time, however, $\coth\eta$ appears rather than $\tanh\eta$, so the first-order correction is more than merely simplified, it vanishes:
\begin{equation} \label{pzExpanded}
    p_\mathrm{z}\approx p_{\mathrm{T}} \sinh\eta \left[1 + \frac{1}{2} \left(\frac{m}{E}\right)^4 \,\frac{\cos\theta}{\sin^4\theta} \right].
\end{equation}

In using eqs \eqref{Eexpanded} and \eqref{pzExpanded} in the following, one eventually replaces $\cos\theta$ with $\tanh\eta$ and $\sin\theta$ with $\sech\eta$ (viz. eq. 5).

\vspace{0.3cm}
\subsection{Corrections of Order $(m/E)^2$}
\subsubsection{Two Particles}
\noindent For two particles, the derivation of eq. \eqref{pairwiseM} involves the expression
\vspace{-0.15cm}
\begin{multline} \label{Edifp}
    (E_1+E_2)^2 - (p_{\mathrm{z}1}+p_{\mathrm{z}2})^2 = (p_{\mathrm{T}1}\cosh\eta_1 + p_{\mathrm{T}2} \cosh\eta_2)^2 \\ - (p_{\mathrm{T}1} \sinh\eta_1 + p_{\mathrm{T}2}\sinh\eta_2)^2
\end{multline}
\noindent which is correct only to zeroth order. Inserting instead eqs \eqref{Eexpanded} and \eqref{pzExpanded} on the left-hand side and retaining only the terms that are of order $(m/E)^2$ gives, since $p_\mathrm{z}$ has no correction of this order,
\vspace{-0.15cm}
\begin{equation*} \begin{split}
    (E_1 + &E_2)^2 - (p_{\mathrm{z}1} + p_{\mathrm{z}2})^2 \approx \left\{p_{\mathrm{T}1} \cosh\eta_1 \left[1 + \frac{1}{2}\left(\frac{m_1}{E_1}\right)^2 \right]\right. \\[0.10cm]
    &\left.\mspace{66mu} \mathrel{+} p_{\mathrm{T}2} \cosh\eta_2 \left[1 + \frac{1}{2}\left(\frac{m_2}{E_2}\right)^2 \right] \right\}^2 - \left( p_{\mathrm{T}1} \sinh\eta_1 + p_{\mathrm{T}2}\sinh\eta_2 \right)^2 \\[0.20cm]
    &\approx p_{\mathrm{T}1}^2 + p_{\mathrm{T}2}^2 + 2p_{\mathrm{T}1}p_{\mathrm{T}2} \cosh(\eta_1-\eta_2) + \left( p_{\mathrm{T}1}\cosh\eta_1 + p_{\mathrm{T}2}\cosh\eta_2  \right) \\[0.10cm]
    &\mspace{200mu} \times \left[p_{\mathrm{T}1}\cosh\eta_1 \left(\frac{m_1}{E_1}\right)^2 + p_{\mathrm{T}2}\cosh\eta_2 \left(\frac{m_2}{E_2} \right)^2\right].
\end{split} \end{equation*}
\noindent Expressions for $p_\mathrm{T}$ and $\phi$ are unaffected by the neglect of $m$, so eq. \eqref{pxySq} can be used as it stands for insertion into eq. \eqref{m12}. It is convenient to write the resulting expression for two-particle invariant mass as
\begin{equation} \label{Dm12def}
    m^2_{12} \approx m^2_{0,12} + m^2_{1,12} \,,
\end{equation}
\noindent where $m^2_{0,12}$ is the zeroth-order expression \textemdash\ that is, eq. \eqref{pairwiseM} \textemdash\ and $m^2_{1,12}$ is the leading-order correction, which factorises as
\vspace{-0.15cm}
\begin{multline} \label{firstOm12}
    m^2_{1,12} = \left(p_{\mathrm{T}1}\cosh\eta_1 + p_{\mathrm{T}2} \cosh\eta_2  \right) \\[0.10cm]
    \times \left[p_{\mathrm{T}1}\cosh\eta_1\left(\frac{m_1}{E_1}\right)^2 + p_{\mathrm{T}2}\cosh\eta_2 \left(\frac{m_2}{E_2}\right)^2\right].
\end{multline}

\subsubsection{Three Particles}
\noindent Corrections to the three-particle expression are derived in the same way. To order $(m/E)^2$,
\vspace{-0.20cm}
\begin{equation*} \begin{split}
    (&E_1 + E_2 + E_3)^2 - (p_{\mathrm{z}1} + p_{\mathrm{z}2} + p_{\mathrm{z}3})^2 = \left\{p_{\mathrm{T}1} \cosh\eta_1 \left[1 + \frac{1}{2}\left(\frac{m_1}{E_1}\right)^2 \right]\right. \\
    &\left.\mspace{100mu} \mathrel{+} p_{\mathrm{T}2} \cosh\eta_2 \left[1 + \frac{1}{2}\left(\frac{m_2}{E_2}\right)^2 \right] + p_{\mathrm{T}3} \cosh\eta_3 \left[1 + \frac{1}{2}\left(\frac{m_3}{E_3}\right)^2 \right] \right\}^2 \\[0.10cm]
    &\mspace{100mu} \mathrel{-} \left(p_{\mathrm{T}1} \sinh\eta_1 + p_{\mathrm{T}2}\sinh\eta_2 + p_{\mathrm{T}3} \sinh\eta_3 \right)^2\\[0.15cm]
\end{split} \end{equation*}
\begin{equation*} \begin{split}
    &\approx p_{\mathrm{T}1}^2 + p_{\mathrm{T}2}^2 + p_{\mathrm{T}3}^2 + 2p_{\mathrm{T}1} p_{\mathrm{T}2}\cosh(\eta_1-\eta_2) + 2p_{\mathrm{T}1} p_{\mathrm{T}3}\cosh(\eta_1-\eta_3)  \\[0.15cm]
    &\mspace{50mu} \mathrel{+} 2p_{\mathrm{T}2} p_{\mathrm{T}3}\cosh(\eta_2-\eta_3) + \left(p_{\mathrm{T}1}\cosh\eta_1 + p_{\mathrm{T}2}\cosh\eta_2  + p_{\mathrm{T}3}\cosh\eta_3  \right) \\[0.10cm]
    &\mspace{68mu} \times \left[ p_{\mathrm{T}1}\cosh\eta_1\left(\frac{m_1}{E_1} \right)^2 + p_{\mathrm{T}2}\cosh{\eta_2} \left(\frac{m_2}{E_2}\right)^2 + p_{\mathrm{T}3}\cosh{\eta_3} \left(\frac{m_3}{E_3}\right)^2 \right].
\end{split} \end{equation*}
\noindent It is again convenient to write, analogously to eq. \eqref{Dm12def},
\vspace{-0.10cm}
\begin{equation}
    m^2_{123} \approx m^2_{0,123} + m^2_{1,123} \,,
\end{equation}
\noindent where $m^2_{0,123}$ is the zeroth-order result given in eq. \eqref{tripletM}: the sum of squared two-particle invariant masses with the three particles taken pairwise. This form invites one to seek a like expression for $m^2_{1,123}$ as the sum of $m^2_{1,ij}$ values. However, counting terms in the expression for the sum $(E_1 + E_2 + E_3)^2$ reveals a complication: the sum has only nine terms of order $(m/E)^2$, but each $m^2_{1,ij}$ contains four such terms (eq. \ref{firstOm12}), so a sum of three requires twelve terms. The remedy is to add and subtract the three extra terms needed, giving
\vspace{-0.15cm}
\begin{equation} \label{firstOm123}
   m^2_{1,123} = m^2_{1,12} + m^2_{1,13} + m^2_{1,23} -\sum_{i=1}^3 p_{\mathrm{T}i}^2\cosh^2\eta_i\left(\frac{m_i}{E_i} \right)^2.
\end{equation}

The form of eq. \eqref{firstOm123} suggests that the leading correction to the three-particle invariant mass is, in fractional terms, smaller than that for two particles, since the three terms subtracted are positive-definite.

\subsubsection{Four and More Particles}
\noindent Like for three particles, counting correction terms provides a pointer to the result for four particles. Writing as before
\begin{equation}
    m^2_{1234} \approx m^2_{0,1234} + m^2_{1,1234} \,,
\end{equation}
\noindent the zeroth-order expression, eq. \eqref{quadrupletM}, contains six pairwise combinations of the four particles. A sum of six corresponding combinations of eq. \eqref{firstOm12} furnishes 24 correction terms, but $(E_1 + E_2 + E_3 + E_4)^2$ provides only sixteen terms of order $(m/E)^2$. Explicit evaluation shows that the eight additional terms required have analogous form to the three additional terms in eq. \eqref{firstOm123}. The result is
\vspace{-0.10cm}
\begin{multline} \label{firstOm1234}
    m^2_{1,1234} = m^2_{1,12} + m^2_{1,13} + m^2_{1,14} + m^2_{1,23} + m^2_{1,24} + m^2_{1,34} \\
    -2\sum_{i=1}^4 p_{\mathrm{T}i}^2\cosh^2\eta_i\left(\frac{m_i}{E_i} \right)^2.
\end{multline}
\noindent This expression suggests a furtherance of the surmise in the comment following eq. \eqref{firstOm123}: the first-order correction to the four-particle value seems smaller again, fractionally, than that for three particles; for here eight positive-definite terms are subtracted from a sum of six two-particle corrections, compared to three from three in eq. \eqref{firstOm123}.

Comparison of eqs \eqref{firstOm123} and \eqref{firstOm1234} leads to the conjecture that the correction term for the invariant mass of a five-particle system may be
\vspace{-0.10cm}
\begin{multline} \label{firstOm12345}
    m^2_{1,12345} = m^2_{1,12} + m^2_{1,13} + m^2_{1,14} + m^2_{1,15} + m^2_{1,23} + m^2_{1,24} + m^2_{1,25} \\ 
    + m^2_{1,34} + m^2_{1,35} + m^2_{1,45} - 3\sum_{i=1}^5 p_{\mathrm{T}i}^2\cosh^2\eta_i\left(\frac{m_i}{E_i} \right)^2,
\end{multline}
\noindent based on the patterns of the algebra in the derivations of eqs \eqref{firstOm123} and \eqref{firstOm1234} and on counts of correction terms. However, readers requiring the result should check explicitly.

\subsubsection{Sensitivity to Polar Angle}
\noindent  In \S\ref{sec:twoPart} it is surmised that the corrections should show sensitivity to momenta directed nearly parallel to the z axis; that is, that the correction terms should involve $\eta$. This does not refer to the $\cosh\eta$ factors in eqs \eqref{firstOm12}, \eqref{firstOm123}, \eqref{firstOm1234} and \eqref{firstOm12345}; for, although $\cosh\eta\rightarrow e^{|\eta|}$ as $\eta\rightarrow \pm\infty$, these factors are always multiplied by a $p_\mathrm{T}$, all of which go to zero in the same limit. Indeed, it is apparent from eq.~\eqref{Eexpanded} that, to zeroth order \textemdash\, which is sufficient in this context \textemdash\, $p_\mathrm{T} \cosh\eta = E$. Hence the leading-order correction has no explicit dependence on $\eta$.

\subsection{Corrections of Order $(m/E)^4$}
\noindent  Since the leading correction does not show the surmised sensitivity to $\eta$, we look for it in the term of order $(m/E)^4$. Returning to the two-particle result of eq. \eqref{Edifp}, but this time keeping all terms in eqs \eqref{Eexpanded} and \eqref{pzExpanded}:
\vspace{-0.10cm}
\begin{equation*} \begin{split}
    (E_1 + &E_2)^2 - (p_{\mathrm{z}1} + p_{\mathrm{z}2})^2 \\[0.1cm]
    &\approx \left\{p_{\mathrm{T}1} \cosh\eta_1 \left[1 + \frac{m_1^2}{2E_1^2} + \frac {m_1^4}{8E_1^4} \left(3 + \frac{4\cos^3\theta_1}{\sin^4\theta_1} \right)\right]\right. \\[0.15cm]
    &\left.\mspace{80mu} \mathrel{+} p_{\mathrm{T}2} \cosh\eta_2 \left[1 + \frac {m_2^2}{2E_2^2} + \frac{m_2^4}{8E_2^4} \left(3 + \frac{4\cos^3 \theta_2}{\sin^4\theta_2} \right)\right] \right\}^2 \\[0.15cm]
    & - \left[p_{\mathrm{T}1} \sinh\eta_1 \left(1 + \frac{m_1^4}{2E_1^4} \; \frac{\cos\theta_1}{\sin^4\theta_1} \right) + p_{\mathrm{T}2}\sinh\eta_2 \left(1+\frac{m_2^4}{2E_2^4} \;\frac{\cos\theta_2}{\sin^4\theta_2} \right) \right]^2. \\[0.10cm]
\end{split} \end{equation*}
\noindent Extending eq. \eqref{Dm12def} to include corrections $m^2_{2,12}$ of order $(m/E)^4$:
\vspace{-0.15cm}
\begin{equation}
    m^2_{12} \approx m^2_{0,12} + m^2_{1,12} + m^2_{2,12} 
\end{equation}
\noindent and using \eqref{tanhEta} to replace $\theta$ by $\eta$, one finds from the equation above that
\begin{equation} \label{secondOm12} \begin{split}
    m^2_{2,12} &= \left(p_{\mathrm{T}1}\cosh\eta_1 + p_{\mathrm{T}2} \cosh\eta_2 \right) \\[0.05cm]
     &\mspace{80mu} \times \left[\frac{p_{\mathrm{T}1} \cosh\eta_1}{4} \left( \frac{m_1}{E_1}\right)^4 \left(3 + 4\sinh^3\eta_1\cosh\eta_1  \right)\right. \\[0.10cm]
     &\left.\mspace{120mu} \mathrel{+} \frac{p_{\mathrm{T}2} \cosh\eta_2}{4} \left(\frac{m_2}{E_2}\right)^4 \left(3 + 4\sinh^3\eta_2 \cosh\eta_2 \right)\right] \\
    &\mathrel{+}  \left[\frac{p_{\mathrm{T}1} \cosh\eta_1}{2} \left( \frac{m_1}{E_1}\right)^2 + \frac{p_{\mathrm{T}2} \cosh\eta_2}{2} \left(\frac{m_2}{E_2}\right)^2 \right]^2 \\
    &\mathrel{-}\left(p_{\mathrm{T}1}\sinh\eta_1 + p_{\mathrm{T}2} \sinh\eta_2 \right) \left[p_{\mathrm{T}1}\sinh^2\eta_1\cosh^3\eta_1 \left( \frac{m_1}{E_1} \right)^4\right. \\
    &\left.\mspace{250mu}\mathrel{+} p_{\mathrm{T}2}\sinh^2\eta_2 \cosh^3\eta_2\left(\frac{m_2}{E_2}\right)^4 \right].
\end{split} \end{equation}
\noindent At first sight, eq. \eqref{secondOm12} has explicit $\eta$ dependence in the terms where the combined power of $\sinh\eta$ and $\cosh\eta$ exceeds the power of $p_{\mathrm{T}}$, but the following analysis suggests that this is somewhat illusory.

Momenta directed nearly parallel to the z axis correspond to $\eta\rightarrow \pm\infty$, in which limit $\cosh\eta\rightarrow e^{|\eta|}$ and $\sinh\eta \rightarrow\sgn\eta\, e^{|\eta|}$. In this limit, $p_{\mathrm{T}}\cosh\eta\approx~E$ and $p_{\mathrm{T}} \sinh\eta\approx\sgn\eta\, E$.  Applying this to both particles, eq. \eqref{secondOm12} approaches
\vspace{-0.10cm}
\begin{equation} \begin{split}
    m^2_{2,12} &\rightarrow \left(E_1 + E_2 \right) \left[\sgn\eta_1 \: E_1 \, e^{4|\eta_1|} \left(\frac{m_1}{E_1}\right)^4 + \sgn\eta_2 \: E_2 \, e^{4|\eta_2|} \left(\frac{m_2}{E_2}\right)^4 \right] \\
    &\mspace{24mu}\mathrel{-} \left(\sgn\eta_1 \: E_1 + \sgn\eta_2 \: E_2  \right)\left[ E_1 \,e^{4|\eta_1|} \left( \frac{m_1}{E_1} \right)^4 \right. \\[-0.25cm]
    &\left.\mspace{340mu}\mathrel{+} E_2\,e^{4|\eta_2|} \left( \frac{m_2}{E_2}\right)^4 \right].
    \end{split}
\end{equation}
This is plainly zero if both particles are travelling in the same direction (i.e. when $\sgn\eta_1 = \sgn\eta_2$); remarkably, it is also zero if they are travelling in opposite directions. That is, the looked-for $\eta$ dependence is not apparent where it might be expected to be largest, namely at large $\eta$.

The opposite limit is $\eta\rightarrow 0$, in which $\cosh\eta \rightarrow 1$ and $\sinh\eta\rightarrow 0$. Applying this limit to both particles, one finds that eq. \eqref{secondOm12} approaches
\begin{equation} \begin{split}
    m^2_{2,12} \rightarrow\: p^2_{\mathrm{T}1} \left(\frac{m_1}{E_1}\right)^4 +\, p^2_{\mathrm{T}2} \left(\frac{m_2}{E_2} \right)^4 &+ \frac{3p_{\mathrm{T}1}p_{\mathrm{T}2}}{4}\left[\left(\frac{m_1}{E_1}\right)^4 + \left(\frac{m_2}{E_2}\right)^4 \right] \\[0.05cm]
    &\mathrel{+} \frac{p_{\mathrm{T}1}p_{\mathrm{T}2}}{2}\left( \frac{m_1}{E_1}\right)^2 \left(\frac{m_2}{E_2}\right)^2.
\end{split} \end{equation}
\noindent This is clearly not zero. It therefore amounts to an identification of $\eta$ dependence, since the previous paragraph shows that, in this same case of the limit applying to both particles, $m^2_{2,12}\rightarrow 0$ as $\eta\rightarrow \pm\infty$.

The extension of this analysis to three and more particles is left as an exercise for the reader who needs it (or is curious).

\section{Summary and Conclusion}
\noindent Section \ref{sec:basics} of this note recapitulates the standard derivations of expressions for the invariant mass of two-, three- and four-particle systems in terms of the measurable quantities $p_{\mathrm{T}}$, $\eta$ and $\phi$. The two-particle expression is widely known; an internet search for the three- and four-particle expressions failed to find any result, suggesting that these are not as well known as their simplicity warrants.

The standard derivations make the assumptions $E \gg m$ and $p_\mathrm{T} \gg m$. Section \ref{sec:corr} explores corrections to these assumptions by expanding expressions in Taylor series in $m/E$, finding that the assumptions are more robust than might be naively expected:
\begin{itemize}
    \item{The leading correction is of order $(m/E)^2$, rather than linear in $m/E$, and the next-to-leading correction is of order $(m/E)^4$.}
    \item{The expressions furnish two sources of corrections which partly cancel to leading order, rendering the net correction for two particles smaller in magnitude that would be the case if either source acted alone.}
    \item{The leading correction for three particles is smaller, in fractional terms, than that for two particles, and those for four and more particles smaller again.}
    \item{The assumption $p_\mathrm{T} \gg m$ suggests that the corrections should show explicit $\eta$ dependence. There is none to leading order. At next-to-leading order, the $\eta$ dependence is modest; in particular, it goes to zero in the situation where it might naively be expected to be largest, namely as $\eta\rightarrow\pm\infty$, i.e. when the momenta of the particles in a collider experiment are nearly aligned to the beam direction.}
\end{itemize}

The centre-of-mass energy delivered by the Large Hadron Collider is so much greater than the mass of any known particle that the analysis of \S\ref{sec:corr} is entirely irrelevant to the practical analysis of experimental data. Nevertheless it it interesting to find that the assumptions underlying the standard derivation are robust: the corrections are smaller than one would expect at first sight.


\begin{thebibliography}{00000}
\bibitem[Wikipedia (2025)]{Wi25}
Wikipedia, \url{https://en.wikipedia.org/wiki/Invariant_mass} (accessed 3 Nov 2025). It is the author's intention to edit this page following release of the present report.
\bibitem[Buckley et al. (2021)]{Bu21}
A.~Buckley, C.~White \& M.~White, \emph{Practical Collider Physics}, Bristol, IOP Publishing, 2021.
\bibitem[Fidalgo (2024)]{Fi24}
Google Colab, \url{https://colab.research.google.com/github/GuillermoFidalgo/Python-for-STEM-Teachers-Workshop/blob/master/notebooks/5-Calculate_invariant_mass.ipynb}, section ``Performing the calculation" (accessed 6 Feb 2025).
\bibitem[Didomeni (2018]{Di18}
University of Rome I, \url{https://www.roma1.infn.it/~didomeni/MEPP/MEPP1819/MEPP1819_add_lez22_1.pdf}, slide 14 (accessed 6 Feb 2025).
\bibitem[Fanti (2012)]{Fa12}
M. Fanti, \url{https://www0.mi.infn.it/~fanti/Particelle4/fanti_PhysicsAtLHC .pdf}, slide 6 (accessed 6 Feb 2025).
\end{thebibliography}
\end{document}